\documentstyle{mn}
\title[HR2875: Spectroscopic discovery of the first  
   B star$+$white dwarf binary]
{HR2875: Spectroscopic discovery of the first  
   B star$+$white dwarf binary}

\author[M.\,R. Burleigh and M.\,A. Barstow]
{Matt Burleigh and Martin Barstow\\
$^1$ Department of Physics and Astronomy, University 
of Leicester, University Rd., Leicester, LE1 7RH \\
}
\date{December 21st 1997}
\def\TD-1{\it TD-1\rm }
\def\tkev{\thinspace{ke\kern-.15em V}}
\def\tev{\thinspace{e\kern-.15em V}}

\begin{document}
\maketitle

\begin{abstract}

We report the discovery, in an Extreme Ultraviolet Explorer (EUVE) short
wavelength spectrum, of an unresolved  
hot white dwarf companion to the 5th-magnitude 
B5Vp star HR2875. This is the first time that a non-interacting 
white dwarf$+$ B star binary has been discovered; previously, the 
the earliest type star known with a white dwarf companion was Sirius
(A1V). Since the white dwarf must have evolved from a main sequence
progenitor with a mass greater than that of a B5V star
($\geq$6.0M$_\odot$), this places a lower limit on the maximum mass for 
white dwarf progenitors, with important implications for our knowledge of
the initial-final mass relation. 
Assuming a pure-hydrogen atmospheric composition, 
we constrain the temperature of the white dwarf to be between 39,000K
and 49,000K. We also argue that this degenerate star is likely to have
a mass significantly greater than the mean mass for white dwarf stars
($\approx$0.55M$_\odot$). Finally, we suggest that other bright B stars (e.g.\
$\theta$ Hya) detected in the extreme ultraviolet surveys of the ROSAT
Wide Field Camera and EUVE may also be hiding hot white dwarf companions. 

\end{abstract}

\begin{keywords} Stars: binaries  -- Stars: white dwarfs
-- Stars: early-type -- Stars: individual: HR2875 -- X-ray: stars.
\end{keywords}

\section {Introduction} 

The extreme ultraviolet (EUV) surveys of the ROSAT Wide Field 
Camera (WFC, Pounds et al. 1993)  and the Extreme Ultraviolet Explorer 
(EUVE, Bowyer et al. 1994) have found a substantial number of white dwarfs, 
in excess of 120. Most of these stars are isolated, but over 30 are now 
known to lie in binary systems (Burleigh 1997). In particular, nearly 20 
unresolved pairs consisting of a hot white dwarf and a bright, 
normal star (spectral type K or earlier) have been found (e.g. Barstow et al.
1994, Vennes et al. 1995, Burleigh et al. 1997, Burleigh 1998). 
Prior to the two 
EUV surveys, these systems were all but unidentifiable, since the normal 
stellar companion completely swamps the optical flux coming from the white 
dwarf. In each case, however, the detection of EUV radiation with the 
spectral signature of a hot white dwarf gave a clue to the existence of the 
previously invisible, faint degenerate companion. Far-ultraviolet spectra 
taken with the International Ultraviolet Explorer satellite (IUE) were then 
used to confirm the identifications. This technique has proved excellent 
for finding these systems in all cases where the normal star is of
spectral type $\sim$A5 or later.

In fact, the earliest type star so far identified by this method to have  
an unresolved hot 
white dwarf companion is Beta Crateris (A2IV$+$WD, Fleming et al. 1991).
Indeed, it took a very careful, detailed, analysis by Barstow et al. (1994) to 
finally confirm this discovery. Unfortunately, even at far-UV wavelengths, 
stars of spectral types early A, B or O will still completely dominate any 
emission from smaller, fainter companions, rendering them invisible even to 
IUE or HST. Overall, the earliest type star known with a white dwarf companion
remains Sirius (A1V$+$DA). 

The spectral type of the normal star in these binaries gives a lower
limit to the mass of the white dwarf progenitor. 
The value of the maximum
mass feasible for producing a white dwarf, and the form of the
initial-final mass relationship (IMFR), are long-standing astrophysical
problems (e.g. Weidemann 1977). Weidemann (1987) gives the upper limit as
8M$_\odot$ in his semi-empirical IFMR. Recent observations of
four white dwarfs in the young open cluster NGC 2516 (Jeffries 1997),
however, imply that the upper mass limit for white dwarf progenitors is
only 5$-$6M$_\odot$. This value is actually in agreement with current stellar
evolutionary models which include moderate core overshoot, but, clearly,
any observations which can place limits on the maximium white dwarf 
progenitor mass have important implications for our theories and models
of stellar evolution, the birth rate of neutron stars and the predicted
rates of type II Galactic supernovae. 

\begin{table*}
\begin{center}
\caption{X-ray and EUV count rates (counts/ksec)}
\begin{tabular}{llcccccccc}
 &  & WFC & & PSPC & & EUVE & & & \\
ROSAT No. & Name & S1 & S2 & (0.1-0.4keV) & (0.4-2.4keV) & 100\AA
& 200\AA & 400\AA & 600\AA \\
RE J0729$-$388 & HR2875 & 61$\pm$14 & 108$\pm$16 & 496$\pm$48 & 0.0 &
94$\pm$12 & 0.0 & 0.0 & 64$\pm$32 \\
\end {tabular}
\end{center}
\end{table*}

HR2875 ($=$HD59635, $=$y Pup) is one of a small number of bright (V$=$5.41)  B 
stars unexpectedly detected in the ROSAT and EUVE all sky surveys. Hiltner 
et al. (1969) classify it as B5Vp, noting that it is overabundant in Si, 
although in the Michigan Catalog of HD Stars (Houk 1982) it receives a B3V 
classification. In this paper we present an analysis of an EUVE spectrum 
of HD2875 which appears to reveal the presence of a previously hidden hot 
white dwarf. If this detection is real and the system is a true 
binary, then HD2875 is the earliest type star known with a white dwarf 
companion. 

\vspace{-0.3cm}

\section{Detection of EUV radiation from HR2875 in the ROSAT WFC and EUVE
surveys}

The ROSAT WFC EUV and X-ray all-sky surveys were conducted between July
1990 and January 1991; the mission and instruments are described
elsewhere (Tr\"umper 1992, Sims et al. 1990). HR2875 (HD59635, y Pup),
catalogued as B5Vp by SIMBAD, is associated in the WFC Bright Source
Catalogue (Pounds et al. 1993) with the source RE J0729$-$388. 
The count rates given in the revised 2RE Catalogue (Pye et al. 1995),
which was constructed using improved methods for source detection,
background screening, etc., are 61$\pm$14 counts/ksec in the S1 filter
and 108$\pm$16 counts/ksec in S2 (see also Table 1). 
The same source was also detected in 
the Extreme Ultraviolet Explorer (EUVE) all-sky survey, which was carried
out between July 1992 and January 1993; the count rates given in Table 1
are taken from the revised Second EUVE Source Catalog (Bowyer et al. 1996).
Finally, the EUV source is coincident with a ROSAT PSPC soft X-ray
detection, although it is only seen in the lower (or soft, 0.1$-$0.4
keV) band. The PSPC count rate was obtained via the World Wide Web from the 
on-line ROSAT All Sky Survey Bright Source Catalogue maintained by the
Max Planck Institute in Germany (Voges et al. 1997). 

The only stars other than white dwarfs whose photospheric extreme 
ultraviolet  radiation has been detected by EUVE are the bright 
B stars $\beta$ CMa (B1II$-$III, Cassinelli et~al. 1996) 
and $\epsilon$ CMa (B2II, Cohen et al. 1996). The photospheric 
continuum of $\epsilon$ CMa is visible in EUVE spectra down to 
$\sim$300{\AA}, although no continuum flux is seen from $\beta$ CMa below the 
HeI edge at 504{\AA}. These two stars are the brightest 
objects in the sky in the EUVE 600{\AA} filter, and in the 
long wavelength spectrometer ($\sim$300$-$740{\AA}). 
They were detected because they lie in a tunnel 
in the interstellar medium of very low HI column density 
(N$_H$$\approx$1$-$2$\times$10$^{18}$ cm$^{-2}$, Welsh 1991). 
Both stars have strong EUV 
and X-ray emitting winds, and in $\epsilon$ CMa 
in the EUVE short and medium wavelength 
spectrometers emission lines from high ionisation species of iron are seen, 
for example. 

The ROSAT and EUVE count rates for HR2875 (B5Vp) are relatively high, and  
immediately mark this EUV source out as unusual. In the EUVE source 
catalogues (Bowyer et al. 1994, 1996, Malina et al. 1994), it is flagged  
as possibly being due to longer wavelength UV radiation leaking through the 
filters, a known problem for EUVE. This problem does not, however, affect 
the ROSAT WFC S2 survey filter. An upper limit to the expected 
contribution to the S2 count rate 
from HR2875's far-UV flux can be estimated (see the ROSAT User's 
Handbook). Assuming a photospheric temperature of $\approx$20,000K, the 
far-UV leakage  contribution to the S2 flux for HR2875 is just 5 counts/ksec, 
compared to the $>$100 counts/ksec detected. Add the fact that the EUV detection 
is also coincident with a PSPC soft X-ray detection, and it can be safely 
assumed that this source is real. 

The origin of this EUV and X-ray radiation is the subject of this paper. 
Note that HR2875 is 
only detected in the ROSAT PSPC lower energy band (0.1$-$0.4keV). Most active 
stars are seen in both the lower (soft) and upper (hard) bands, 
while only one (rather unusual) 
white dwarf has ever be detected in this energy range (KPD0005$+$5106, Fleming 
et al. 1993). The X-ray/EUV colours and count rate ratios (S2/S1$\approx$2) 
for HR2875 are also 
very similar to many of the hot white dwarfs detected by ROSAT and EUVE,
a point emphasised by Motch et al. (1997), who discounted HR2875 as a new
massive X-ray binary but also suspected that it might be hiding a
non-accreting white dwarf. Burleigh et al. (1997) suggested, therefore, 
that HR2875, like almost 20 other stars in the EUV 
catalogues, may be hiding a hot white dwarf companion. 

\vspace{-0.3cm}

\section{EUVE pointed observation and data reduction}

\begin{figure*}
\vspace*{8cm}
\includegraphics{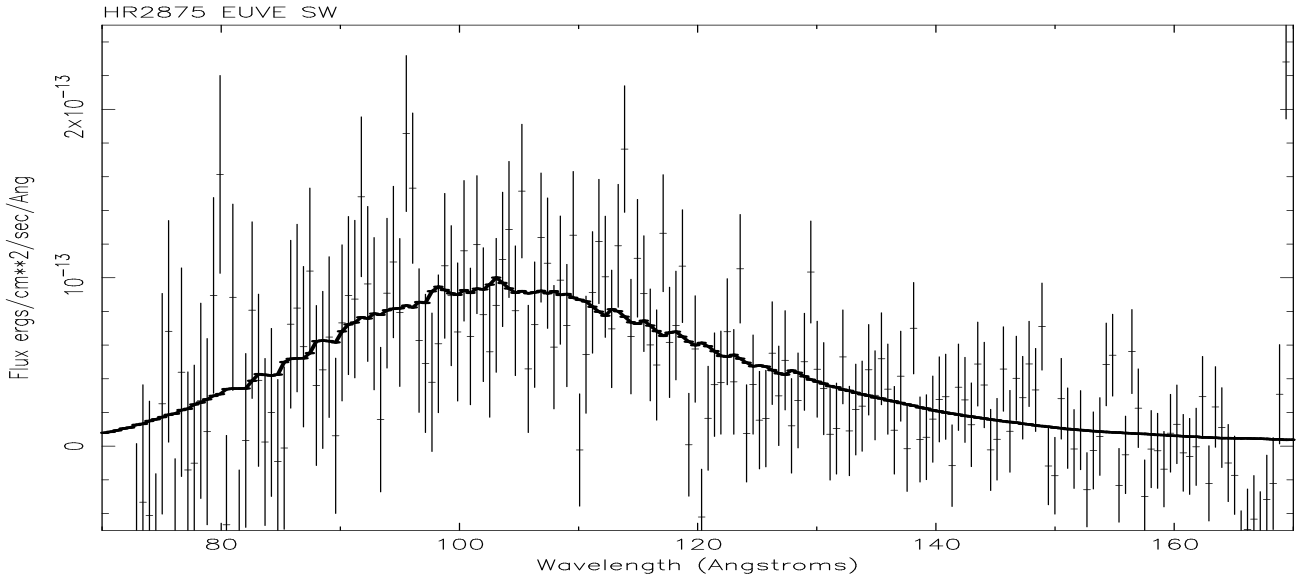}
\caption{\em EUVE short wavelength spectrum of HR2875, binned to the
resolution of the instrument, $\approx$0.5{\AA} (Note that this spectrum was 
further binned by a factor 8 during the analysis). Also shown is a pure
hydrogen white dwarf $+$ ISM model for log $g=$8.5, $T_{\em eff}=$43,400K, 
$N_{HI}$$=$2.1$\times$10$^{19}$atoms cm$^2$,
$N_{HeI}$$=$2.4$\times$10$^{18}$atoms cm$^2$, and
$N_{HeII}$$=$8.9$\times$10$^{17}$atoms cm$^2$. }
\end{figure*}

HR2875 was observed by EUVE in dither mode during April 1996. 
Two separate observations were made (GO458 \& GO459, PI: R.S. Polidan), of
$\approx$85,000 and $\approx$65,000 secs respectively, which became
publicly available one year later. We have extracted the spectra from the 
images ourselves, using standard IRAF procedures. Our general reduction
techniques are described in earlier work (e.g.\ Barstow et al. 1997). 

The target was not detected in either the medium wavelength (140$-$380\AA) 
or long wavelength (280$-$760\AA) 
spectrometers, and the signal/noise of the data in the short wavelength
spectrometer (70$-$190\AA) in both observations was very poor. 
Consequently, we have co-added the two short wavelength spectra to improve the
signal/noise for the subsequent analysis. 
This pointed observation, shown in Figure 1, reveals a weak continuum,
with a peak flux of $\sim$1.5$-$2.0$\times$10$^{-13}$ergs cm$^2$
sec$^{-1}$ {\AA}$^{-1}$,   
characteristic of the many hot white dwarfs observed by EUVE with
this spectrometer, and, in addition, there is no evidence for strong 
emission features. This would appear to rule out a hot wind as the source
of the EUV and soft X-ray emission, as we might have expected to see
emission lines from high ionisation species of e.g\ iron. Similarly, we
can also eliminate the possibility that this might be an RS CVn binary
such as $\chi$ Uma (Schrijver et~al. 1995), or that HR2875
might be hiding an active late-type companion, such as in the B8V$+$K2IV
binary Algol (Stern et~al. 1995). 
In both those systems, high ionisation features of e.g.\
iron, oxygen, nickel and calcium are seen in EUVE short wavlength spectra. 

\vspace{-0.3cm}

\section{Analysis}

Since we suspected that the EUV continuum detected by the short
wavelength spectrometer is being produced by a hot white dwarf, we
decided to try to match this spectrum (together with the ROSAT WFC S1, S2
and PSPC broad band fluxes) 
with a grid of white dwarf$+$ISM model
atmospheres, in order to constrain the possible atmospheric parameters
(temperature and surface gravity) of
the degenerate star and the interstellar column densities of HI, HeI and
HeII. Unfortunately, there are no spectral features (e.g.\ H absorption
lines) in this region of the spectrum to give us an
unambiguous determination of $T$ and log $g$. However, by making a range of
assumptions to reduce the number of free parameters in our models, we can
place constraints on e.g.\ the white dwarf's temperature. 

\begin{table}
\begin{center}
\caption{Hamada-Salpeter zero-temperature mass-radius relation}
\begin{tabular}{ccccc}
log $g$ & $M_{\em WD}$ & $R_{\em WD}$ & $R_{\em WD}$ & ($R_{\em
WD}$/$D$)$^2$ \\
 & M$_\odot$ & R$_\odot$ & $\times$10$^6$m & {\em where D$=$170pc} \\
7.5 & 0.30 & 0.017 & 11.832 & 5.089$\times$10$^{-24}$ \\
8.0 & 0.55 & 0.013 & 9.048 & 2.975$\times$10$^{-24}$ \\
8.5 & 0.83 & 0.009 & 6.264 & 1.426$\times$10$^{-24}$ \\
9.0 & 1.18 & 0.006 & 4.176 & 0.634$\times$10$^{-24}$ \\
\end {tabular}
\end{center}
\end{table}

\begin{table}
\begin{center}
\caption{White dwarf atmospheric parameters and interstellar column
densities}
\begin{tabular}{ccccc}
log $g$ & $T_{\em eff}$ (K) & $N_{HI}$ $\times$10$^{19}$ & $N_{HeI}$ 
& $N_{HeII}$ \\
 & \& 90\% range & \& 90\% range  & $\times$10$^{18}$ & 
$\times$10$^{18}$ \\
7.5 & 40,500 (39,200$-$41,700) & 2.8 (2.3$-$3.3) & 3.1 & 1.2 \\
8.0 & 41,000 (40,100$-$42,300) & 2.4 (2.0$-$2.9) & 2.7 & 1.0 \\
8.5 & 43,400 (41,900$-$45,200) & 2.1 (1.7$-$2.6) & 2.4 & 0.9 \\
9.0 & 46,800 (45,500$-$48,400) & 1.9 (1.5$-$2.3) & 2.1 & 0.8 \\
\end {tabular}
\end{center}
\vspace{-0.5cm}
\end{table}

Firstly, we assume that the white dwarf has a pure-hydrogen atmosphere. This is
a reasonable assumption to make, since Barstow 
et al. (1993) first showed that for $T_{\em eff}$$<$40,000K hot white dwarfs have
essentially pure-H atmospheres. We can then fit a range of models, each
fixed at a value of the surface gravity log $g$. Before we can do this,
however, we need to know the normalisation parameter of each model, which
is equivalent to (Radius$_{\em WD}$/Distance)$^2$. 
We can use the {\em Hipparcos} parallax 
(5.86$\pm0.51$ milli-arcsecs., ESA 1997)    
to calculate the distance to the system (170$^{+17}_{-13}$
parsecs), and the Hamada-Salpeter 
zero-temperature mass-radius relation to give us the radius of the white
dwarf corresponding to each value of the surface gravity, 
since we have no {\em a priori} knowledge of the star's temperature (see
Table 2).

We can also reduce the number of unknown free parameters in the ISM model. 
From EUVE spectroscopy, 
Barstow et al. (1997) measured the line-of-sight interstellar column
densities of HI, HeI and HeII to a number of hot white dwarfs. They found
that the mean H ionisation fraction in the local ISM was 0.35$\pm$0.1,
and the mean He ionisation fraction was 0.27$\pm$0.04. From these
estimates, and assuming a cosmic H/He abundance, we calculate the ratio 
 $N_{HI}$/$N_{HeI}$ in the local ISM$=$8.9, and
$N_{HeI}$/$N_{HeII}$$=$2.7. We can then fix these column density ratios
in our model, leaving us with just two free parameters - temperature and
the HI column density. 
 
The model fits at a range of surface gravities from log $g=$7.5$-$9.0 are
summarized in Table 3.

\vspace{-0.3cm}

\section{Discussion}

We have discovered an unresolved hot white dwarf companion to the 5th
magnitude B star HR2875 (y Pup). This is the first time a hot white dwarf
$+$ B star binary has been detected, and it has important 
implications for our understanding of white dwarf and stellar evolution,
since a white dwarf companion to such an early-type star must have
evolved from a very massive progenitor, close to the upper limit for
white dwarf formation. According to Lang (1992) a B5V star has a mass of
5.9M$_\odot$ (or 6.5M$_\odot$ according to Allen, 1973), and if the
spectral type is as early as B3V (as classified in the Michigan Catalog
of HD Stars, Houk 1982) then it will of course have a slightly higher 
mass. This is $\geq$5$-$6M$_\odot$ upper mass limit for white dwarf
progenitors of Jeffries (1997), but still significantly less than the
8M$_\odot$ upper limit from Weidemann (1987). 

Since it must have evolved from such a massive progenitor, it is likely
that this white dwarf also has a higher mass than the mean for these
degenerates (0.583$\pm$0.078M$_\odot$, Marsh et al. 1997).  
We can use the theoretical initial-final mass relation between 
main sequence stars and white dwarfs of Wood (1992), to calculate the mass of
the white dwarf if 
its progenitor was only slightly more massive than HR2875: 

\vspace{0.35cm}

$M_{\em WD}$$=$Aexp(B$\times$$M_{\em MS}$) 

\vspace{0.35cm}

where~A$=$0.49462M$_\odot$ and
B$=$0.09468M$_\odot$$^{-1}$. For
$M_{\em MS}$$=$6.5M$_\odot$, we find $M_{\em WD}$$=$0.91M$_\odot$.
This would suggest the surface gravity of the white dwarf log $g$$>$8.5.
In the log $g=$8.5 model, $T_{\em eff}=$43,400K and the white dwarf has a
V magnitude $=$16.4 (calculated from the model flux at 5500{\AA}). 

HR2875 is not known to display radial velocity variations, and {\em
Hipparcos} found no evidence for micro-variations in its proper motion
across the sky to suggest that this might be a relatively short period
binary system (P$<$few years, ESA 1997). 
However, it is clearly important to study
this system further, and if radial velocity variations are detected then
the binary parameters and the white dwarf's mass can be constrained, with
important implications for our knowledge of the initial-final mass
relation. Additionally, further study of the B star primary might reveal
evidence for past interaction. Could the over-abundance of Si detected in
this object by Hiltner et al. (1969) be due to accretion from the wind of
the evolved giant progenitor to the white dwarf, as in the WD$+$K2V
binary RE J0357$+$28 (Jeffries, Burleigh and Robb 1996)? 

Are there more early-type star$+$hot white dwarf binaries in the ROSAT
and EUVE catalogues awaiting discovery? From its ROSAT WFC EUV and PSPC
soft X-ray count rates and colours (S1$=$52$\pm$7 counts/ksec, S2$=$148$\pm$12
counts/ksec, PSPC$=$124$\pm$24 counts/ksec - all in the lower band), 
we strongly suspect that the 4th magnitude star 
$\theta$ Hya (B9.5V, $=$HR3665, $=$RE J0914$+$02) also
hides a hot white dwarf companion, and indeed EUVE is scheduled to
observe this star spectroscopically in March 1998. Other B stars detected
in the EUV surveys may also have non-interacting degenerate companions,
e.g.\ ADS10129C (B9.5V, $=$HD150100, $=$RE J1636$+$52) which also has EUV
and soft X-ray count rates very similar to known hot white dwarfs 
(ROSAT WFC 
S1$=$12$\pm$4 counts/ksec, S2$=$46$\pm$11 counts/ksec, PSPC$=$72$\pm$15
counts/ksec - again, all in the lower band). Unfortunately these objects are
probably too faint to be detected by the EUVE spectrometers.   

\vspace{-0.3cm}

\section*{Acknowledgements}

Matt Burleigh and Martin Barstow acknowledge the support of PPARC, UK.
We thank Detlev Koester (Kiel) for the use of his white dwarf model
atmosphere grids.

{\it Note added in proof} After this paper had been submitted for
publication, we learned of a similar study by Vennes et al. (1997, ApJ,
491, L85). Their conclusions about the properties of the white dwarf are
in agreement with ours.

{}


\vspace{-0.3cm}

\begin{thebibliography}{}

\bibitem{} Allen C., 1973, 
{\em Astrophysical Quantities}, Athlone Press, London

\bibitem{} Barstow M.A., et~al., 1993, MNRAS, 264, 16

\bibitem{}
Barstow M.A., Holberg J.B., Fleming T.A., Marsh M.C., Koester D.,
Wonnacott D., 1994, MNRAS, 270, 499

\bibitem{} Barstow M.A., Dobbie P.D., Holberg J.B., Hubeny I., Lanz T.,
1997, MNRAS, 286, 58

\bibitem{} Bowyer S., Lieu R., Lampton M., Lewis J., Wu X., Drake J.J.,
Malina R.F., 1994, ApJS 93, 569

\bibitem{}
Bowyer S., Lampton M., Lewis J., Wu X., Jelinsky P., Malina R.F., 
1996, ApJS, 102, 129

\bibitem{}
Burleigh M.R., 1997, {\em PhD. Thesis}, University of Leicester

\bibitem{}
Burleigh M.R., 1998, in {\em Ultraviolet Astrophysics$-$Beyond the IUE
Final Archive}, ESA Publication SP-413, in press

\bibitem{}
Burleigh M.R., Barstow M.A., Fleming T.A., 1997, MNRAS, 287, 381

\bibitem{}
Cassinelli J.P., et~al., 1996, ApJ, 460, 949

\bibitem{}
Cohen D.H., Cooper R.G., MacFarlane J.J., Owocki S.P., Cassinalli J.P.,
Wang P., 1996, ApJ, 460, 506

\bibitem{} ESA, 1997, {\it The Hipparcos Catalogue}, ESA SP$-$1200

\bibitem{} Fleming T.A., Schmitt J.H.M.M., Barstow M.A., Mittaz J.P.D., 
1991, A\&A, 246, L47

\bibitem{} 
Fleming T.A., Werner K., Barstow M.A., 1993, ApJ, 416, L79

\bibitem{} 
Hamada T., Salpeter E.E., 1961, ApJ, 134, 683

\bibitem{} 
Hiltner W.A., Garrixon R.F., Schild R.E., 1969, ApJ, 157, 313

\bibitem{} Houk N., 1982, {\em Michigan Spectral Survey}, University of
Michigan

\bibitem{}
Jeffries R.D., Burleigh M.R., Robb R.M., 1996, A\&A, 305, L45

\bibitem{}
Jeffries R.D., 1997, MNRAS, 288, 585

\bibitem{} Lang K.R., 1992, {\em Astrophysical Data}, Springer-Verlag 
(New York)

\bibitem{}
Malina R.F., et~al., 1994, AJ, 107, 751

\bibitem{} Marsh M.C., et~al., 1997, MNRAS, 286, 369

\bibitem{} 
Motch C., Haberl F., Dennerl K., Pakull M., Janot-Pacheco E., 1997,
A\&A, 323, 853

\bibitem{} Pounds K.A., et~al., 1993, MNRAS, 260, 77

\bibitem{} Pye J.P., et~al., 1995, MNRAS, 274, 1165

\bibitem{} Schrijver C.J., Mewe R., Van Den Oord G.H.J., Kaastra J.S.,
1995, A\&A, 302, 438

\bibitem{} Sims M.R., et~al., 1990, Opt. Eng., 29, 649

\bibitem{} Stern R.A., Lemen J.R., Schmitt J.H.M.M., Pye J.P., 1995, ApJ,
444, L45

\bibitem{} Tr\"umper J., 1992, QJRAS, 33, 165

\bibitem{} Vennes S., Mathioudakis M., Doyle J.G., Thorstensen J.R.,
Byrne P.B., 1995, A\&A, 299, L29 

\bibitem{} Voges W., et~al., 1997,  A\&A, in press

\bibitem{} Weidemann V., 1977, A\&A, 59, 411

\bibitem{} Weidemann V., 1987, A\&A, 188, 74

\bibitem{} Welsh B.Y., 1991, ApJ, 373, 556

\bibitem{} Wood M.A., 1992, 386, 539

\end{thebibliography}
\end{document}